\documentclass[aps,prl,twocolumn,showpacs,amsmath,amssymb,superscriptaddress,a4paper,10pt]{revtex4-1}
\usepackage{graphicx}
\usepackage{color}
\usepackage{CJK}

\newcommand{\Yxuv}{Y_{XUV}}

\begin{document}
\begin{CJK*}{UTF8}{gbsn}

\title{Real-time probing electron dynamics of an atom in a strong infrared laser field}

\author{Yunpei Deng (邓蕴沛)}
\email{deng@fhi-berlin.mpg.de}
\affiliation{Fritz-Haber-Institut der Max-Planck-Gesellschaft, Faradayweg 4-6, 14195 Berlin, Germany}
\author{Xinhua Xie (谢新华)}
\email{xinhua.xie@tuwien.ac.at}
\affiliation{Photonics Institute, Vienna University of Technology, Gusshausstrasse 27, A-1040 Vienna, Austria}

\date{\today}

\begin{abstract}
We present theoretical studies on real-time probing the electron density evolution of an atom in a strong infrared (IR) laser field with few-cycle near-infrared (NIR) and attosecond extreme-ultraviolet (XUV) pulses. Our results indicate that the electron density near the tunneling barrier is reflected in the additional tunneling ionization yield with a delayed NIR pulse and the electron density near the nucleus can be probed by the single photoionization yield with a delayed XUV pulse. It turns out the NIR-probing scheme can be used to study the polarization of the system in an external IR field and the XUV-probing can be additionally applied to explore excitation dynamics during and after the IR field interaction.
\end{abstract}

\maketitle
\end{CJK*}



Strong field physics, which describes the strong laser field interaction with atoms and molecules, has become a fascinating research direction, especially along the generation of attosecond pulses and their broad applications\cite{Scrinzi06,krausz09}.
Strong field phenomena in general involve electron ionization or excitation processes which happen on a very short, femtosecond or attosecond, timescale.
There are various probing techniques to time-resolve such ultrafast processes. 
Since a decade ago, XUV-pump-NIR-probe method has been applied to measure Auger decay process and ionization dynamics of noble gases with subfemtosecond resolution\cite{drescher:auger,uiber07}.
Later attosecond XUV transient absorption spectroscopy has been applied to ultrafast science, e.g. measuring the movement of valence electrons\cite{Gulielmakis2010}, probing time-dependent molecular dipoles \cite{neidel13} and investigating autoionization dynamics of atoms\cite{wang10}.
Besides, there are several other probing methods with attosecond temporal resolution but without using XUV attosecond pulses, such as photoelectron spectroscopy\cite{huismans11,xie12,Boguslavskiy2012} and high-harmonics spectroscopy\cite{olga09:co2,shiner11,worner11}.

With the development of ultrafast lasers and the applications of optical parametric amplification technique, ultrafast strong field physics is extended from the Ti:Sa based near-infrared (NIR) wavelength to infrared (IR) regime with wavelength up to several micrometers\cite{Andriukaitis11,Deng12,Lanin13,Hong14,Mayer13}.
IR lasers are versatile sources which already show their advantages on the generation of attosecond pulses based on high harmonic generation\cite{Marcus12,Popmintchev12,gong12,Chen14}, controlling molecule dissociation\cite{Znakovskaya12,Jia14} and studies on filamentation\cite{Kartashov12}.
In strong field physics, the transition from the multiphoton to the tunneling regime for ionization is characterized by the Keldysh parameter\cite{keldysh64}, which scales quadratically with the laser wavelength.
The longer wavelength not only moves the ionization deeper into the tunneling regime, but furthermore allows for the control of the molecular dissociation at larger internuclear distances.
With IR laser pulse, the increased spacing between successive half-cycles gives a sufficient time window to capture the dynamics of an sub-femtosecond relaxation process, before the identical process is re-triggered by the next half-cycle of the laser pulse.
Therefore, it provides an ideal tool for field triggered pump-probe experiments.

\begin{figure}[ht]
\centering
\includegraphics[width=0.45\textwidth,angle=0]{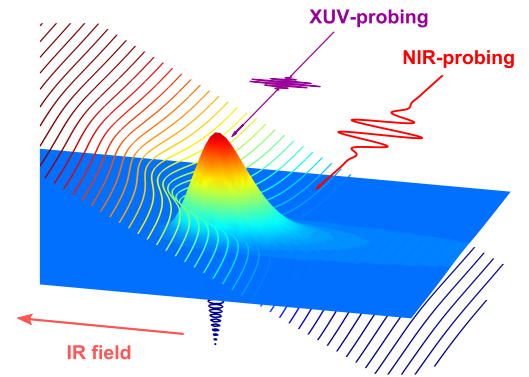}
\caption{Schematic view of XUV- and NIR-probing of a strong IR field driven system.
The density distribution of a IR field distorted ground state wave function is presented as the color coded surface.
The contour lines indicates the tilted Coulomb potential due to the external IR field.}\label{fig:peda}
\end{figure}

In this letter, we report theoretical studies on real-time probing of the electron density evolution of an atom in a IR field with few-cycle NIR and attosecond XUV probe pulses.
We use ionization yield, a measurable quantity in experiments, as the study object from two-color pump-probe simulations.
In the case of NIR-probing, tunneling ionization dominates the additional ionization, while in the case of XUV-probing, single-photon ionization dominates the additional ionization because of low pulse intensity and high photon energy.
We find the probing quantities follow the electron density near the tunneling barrier for the NIR-probing case and the electron density near the origin of the system for the XUV-probing case, as illustrated in Fig.~\ref{fig:peda}.
Because of probing on different quantities, we can get access to the polarization of the electron wave function and the excitation of the system during field interaction of the system with a strong IR field.

\begin{figure}[ht]
\centering
\includegraphics[width=0.45\textwidth]{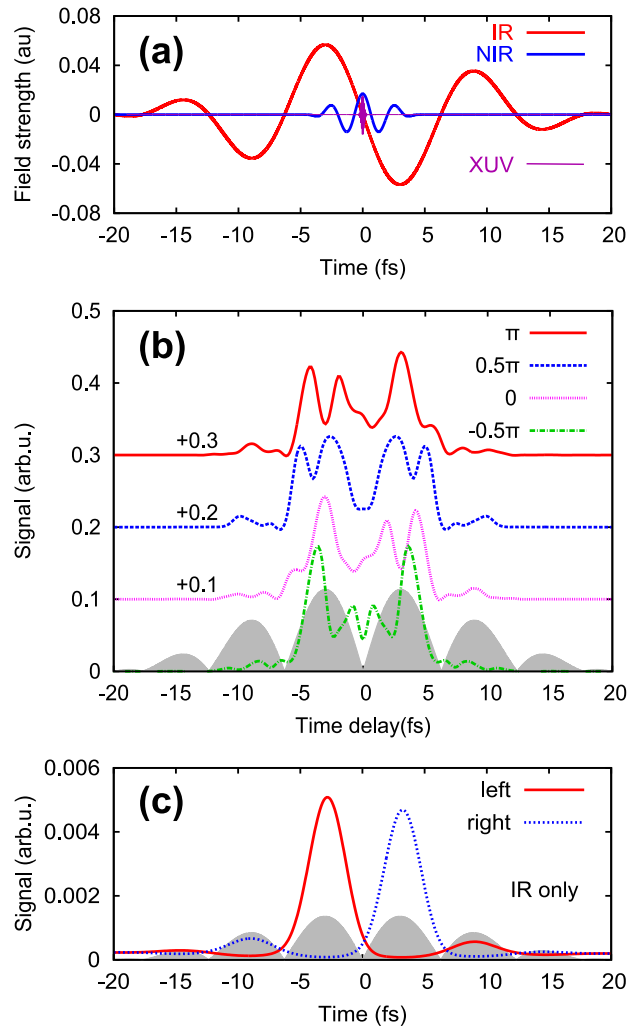}
\caption{(a) Electric fields of the IR(CEP=0.5$\pi$), NIR (CEP=0) and XUV pulses used in the simulations.
(b) Additional ionization yield as a function of the delay time between NIR and IR pulses for four different CEPs of NIR pulses. Additional amount of values are added to the signals to have a better comparison.
(c) Electron density at left (at position (-8,0)) and right (at position(8,0)) side of the atom over wave function evolution time.
The filled gray lines in (b) and (c) indicate the absolute value of the IR electric field over time.}\label{fig:nir-ir}
\end{figure}

We solve the time-dependent Schr\"{o}dinger equation with the single-active-electron approximation in a two-dimensional Cartesian coordinate with the pseudospectral method within a spatial box [-120, 120] a.u. in both directions.
A step-size-adapted propagator is employed with the Runge-Kutta method to control the numerical errors during the electron wave function evolution in the external field.
More details of our simulation methods can be found in \cite{xie07jmo,xie07pra}.
In the simulations, we use a hydrogen-like atomic system which has a ground state energy -0.5 a.u. with a screened Coulomb potential $V(x,y)=-1/\sqrt{x^2+y^2+0.64}$.
Linearly polarized electric fields along the x-coordinate are applied in the simulations.
The vector potential $A$ of the laser fields are defined as $A_F(t)=A_{F0}(t-t_F)\sin{(\omega_F(t-t_F)+\phi_F)}$ with
F=XUV, NIR, IR, where $\omega_F$ and $\phi_F$ are the center frequency and the carrier-envelope phase (CEP), respectively.
$t_F$ is the peak position of the pulse, which is used to control the delays between different pulses.
We ensure $A(-\infty)=A(+\infty)=0$, which excludes any unphysical dc component in the pulses.
In the simulations, sine-square envelopes $A_{F0}(t)=({\mathcal{E}_F}/{\omega _F})\sin^2{({\pi t}/{2\tau_F})}$, are used with full width at half maximum (FWHM) $\tau_F$.
As shown in Fig.~\ref{fig:nir-ir}(a), pulse durations of XUV, NIR and IR are chosen as 150 as, 4 fs and 20 fs with the center wavelength 20 nm, 800 nm and 4 $\mu$m, respectively, as illustrated in the Fig.~\ref{fig:nir-ir}(a).
With such pulse durations, both NIR and IR pulses are few-cycle pulses, and the CEPs of them are critical to the field shapes and important for strong field phenomena\cite{baltuska03:nature,paulus03:phase-measurement,xie12:cep}.
For the IR pulse, we choose CEP=$0.5\pi$ such that the electric field is anti-symmetric around the pulse center.
The IR pumping pulse has a peak intensity of 1.26$\times10^{14}$ W$/$cm$^2$.
With such intensity, the ionization of the model atom is in the tunneling regime according to the Keldysh parameter $\gamma=$0.27\cite{keldysh64}.
The probing XUV and NIR pulses have peak intensities of 1$\times10^{13}$ W$/$cm$^2$, respectively.
To do the pump-probe simulations, we fix the IR pulse and scan the delay between the IR and XUV or NIR by varying $t_F$ of XUV or NIR.


First, we present our results on NIR-probing.
We perform simulations to scan the time delay between the NIR pulse and the IR pulse for four different CEPs ($-0.5\pi$, 0, 0.5$\pi$, $\pi$) of the NIR pulse.
To get the ionization yield, we integrate the electron density by projecting the bound states away from the final electron wave function when the laser field is completely off.
In Fig.~\ref{fig:nir-ir}(b), we show the additional ionization induced by the combined NIR and IR field, which is defined as $\Delta\eta(\tau)=\eta(\tau)_{NIR+IR}-\eta_{NIR}-\eta_{IR}$, where $\eta(\tau)_{NIR+IR}$ is the ionization yield with the superimposed field.
$\eta_{NIR}$ and $\eta_{IR}$ are the ionization yields with only NIR and IR pulse, respectively.
It evidently shows that the extra ionization yields have strong modulation over the time-delay between the NIR and the IR laser fields for all CEPs.
We notice that the change of the yield is always positive, which indicates ionization enhancement due to the overlapping of the two pulses.
First, we focus on the case of CEP=0 (magenta line in the Fig.~\ref{fig:nir-ir}(b)).
The NIR field is a so-called "cosine" pulse, which has one dominate field peak in the center of the pulse.
The modulation of the additional yield follows the intensity of the IR field.
It has a maximal peak at delay time a quarter optical cycle of the IR pulse before 0, where the IR field has its peak field strength.
Intuitively, such modulation depends on the superimposed laser field strength.
The field of the IR field points to the same direction as the NIR peak field at time delay a quarter of the optical cycle of the IR pulse before 0, while on the other side of the IR field, the IR field points to the different direction.
By changing the CEP of the NIR field, the peak strength and the field direction at the time corresponding to the peak field strength changes accordingly.
For CEP=$\pi$, the field shape is the same as CEP=0 but the center peak points to the opposite direction, which leads to a maximal peak of additional ionization yield at time delay a quarter optical cycle after 0.
For the case of CEP=0.5$\pi$ and CEP=-0.5$\pi$, the NIR fields are anti-symmetric which leads to equally strong peaks at positive and negative sides of the time delay.
It suggests that the ionization enhancement is caused by the combined waveform of NIR and IR field.

To further understand the reason for the enhancement, we calculate the electron density near the nucleus when the atom interacts with a IR only laser field as shown in Fig.~\ref{fig:nir-ir}(a).
In Fig.~\ref{fig:nir-ir}(c), we show the electron density at the left and right sides 8 a.u. away from the nucleus over the evolution time.
These positions are below the tunneling barrier when the laser field reaches its peak field strength.
The electron density signal at these two positions follows the electric field of the IR pulse.
On the left side of the nucleus, there are peaks at t=-3.3 fs (a quarter of the IR optical cycle before the pulse center) and t=9 fs.
On the right side of the nucleus, there are peaks at t=3.3 fs (a quarter of the IR optical cycle after the pulse center) and t=-9 fs, which is directly related to the field shape of the IR pulse.
It shows that around the peak of the IR field when the potential barrier is bent by the IR field, the wave function will be polarized along the laser polarization direction, such that electron density will driven to the opposite direction of the laser field, as illustrated in the Fig.~\ref{fig:peda}.
At this particular time, if additional NIR laser field is applied, ionization will be enhanced when the electric field of the NIR field points to the same direction as that of the IR pulse.
By comparing Fig.~\ref{fig:nir-ir}(b) and (c), we notice the extra ionization yield over the time delay follows the electron density changing over time except for the oscillation in the yield due to different CEPs of the NIR pulse.

In an intuitive picture, the IR field induces strong polarization of the bound system and part of the wave function will be moved to the continuum.
The probing NIR field will interact with the polarized bound system and preferentially further tunneling ionizing it from the wave function near the tunneling barrier.



Now, we turn to the case of XUV-probing.
We investigate how the total XUV photoelectron yields vary during ionization by a IR few-cycle laser field.

The XUV ionization yield as a function of time delay $\tau$ between the IR pulse and the XUV pulse
is calculated as \cite{Smirnova06,Smirnova07}
\begin{equation}
\Yxuv(\tau)=|\langle\Psi_{XUV}(\tau)|\Psi_{XUV}(\tau)\rangle|^2 \label{eq:xuv-yield}
\end{equation}
where $\Psi_{XUV}$ is the additional wave function amplitude due to the XUV pulse
\begin{equation}
|\Psi_{XUV}(\tau)\rangle=(1-|\Psi_{IR}\rangle\langle\Psi_{IR}|)|\Psi_{IR+XUV}(\tau)\rangle
\end{equation}
and $\Psi_{IR+XUV}(\tau)$ is the electron wave function calculated with an XUV pulse, which is delayed by $\tau$ with respect to the IR laser pulse, while $\Psi_{IR}$ is the wave function propagated with the only IR laser pulse.
The matrix element in Eq.~\ref{eq:xuv-yield} is evaluated at some time $t>\tau$ after the XUV pulse is over.
The photon energy of the XUV pulse is 62 eV and the intensity 1$\times10^{13}$ W/cm$^2$, which ensures that the XUV induced ionization is dominated by single-photon ionization.
During the interaction of the atom with the IR field, the XUV pulse induces extra ionization which is used as the probe signal.

\begin{figure}[htbp]
\centering
\includegraphics[width=0.45\textwidth,angle=0]{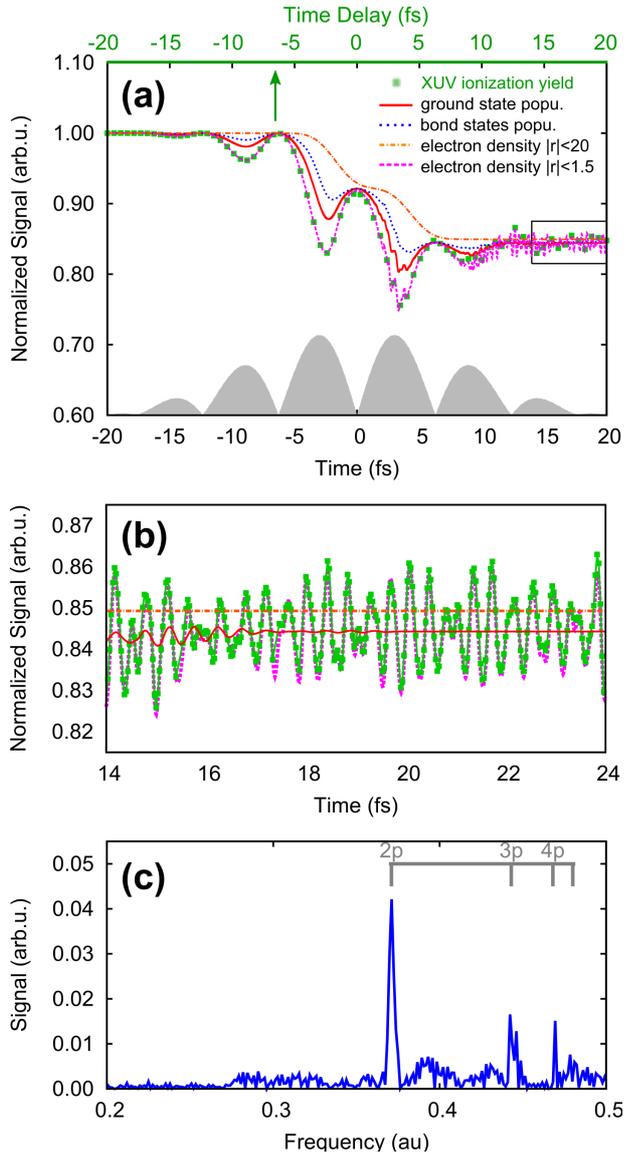}
\caption{(a) The XUV ionization yield is presented as green square points over the time delay.
The population of the ground state and the bound states up to the 4\textit{p} state are shown over time as the red and blue lines, respectively.
The magenta line and the orange line indicate the electron density near nucleus and the overall electron density (with absorption boundary) of the wave function over time.
The gray filled line is the same as those in Fig.~\ref{fig:nir-ir}.
(b) A Zooming in of the rectangular region of panel (a) and extension to larger time delay after the IR pulse to present the fast oscillations.
(c) Fourier transformed spectra of the electron density near the nucleus over time.
The gray bar indicates the beating energy between the 2\textit{p}, 3\textit{p} and 4\textit{p} states and the 1\textit{s} ground states.} \label{fig:xuv-ir}
\end{figure}

The simulated XUV ionization yields as a function of the time delay between the XUV and IR pulse are presented as green square points in the Fig.~\ref{fig:xuv-ir}(a).
The XUV ionization yields are normalized to that with XUV pulse only.
The yields oscillate and gradually decrease as scanning through the IR pulse.
The oscillation follows the electric field strength of the IR pulse.
The yield has local minimum at the time delay around when the IR field has a peak.
Intuitively, the XUV ionization yield should be proportional to the electron density in the bound states from which the photoionization happens\cite{uiber07}.
To compare this quantity with the XUV ionization yield, we include the population of the ground state and the population of bound states from the 1\textit{s} up to the 4\textit{p} state during the IR only interaction with the atom in Fig.\ref{fig:xuv-ir}(a) as the red line and the blue line, respectively.
It is clear that the XUV ionization yield roughly follows the populations of the ground state except that the XUV ionization yield has much stronger modulation.
In another words, the XUV ionization yield is not exactly determined by the population in the bound states.
We do one more simulations with absorption boundaries such that free electrons will be absorbed by the boundaries at a distance 20 a.u. away from the origin and only bounded electrons stays in the calculation box.
The electron density remains in the box then stands for the population of all bounded states, which is presented as the orange line in Fig.~\ref{fig:xuv-ir}(a).
It has a step-wise structure, and at the end of the laser pulse there is small amount of discrepancy between the remaining electron density and the population of the ground state.
It implies that some populations are left in the excited states after the interaction with the IR laser field.
We will further discuss on the excitation effect afterwards.
So far in the discussion, it has not been clear which quantity the XUV ionization yield represents.

By checking the evolution of the electron density of the system in the IR field, we discover that the electron density near the nucleus match very nicely with the XUV ionization yield.
We define the electron density near the nucleus by
\begin{equation}
\label{eq:n0}
N_0(t) =\iint e^{\frac{-1.38(x^2+y^2)}{W^2}}|\Psi_{IR}(x,y;t)|^2 dxdy
\end{equation}
The magenta line in Fig.~\ref{fig:xuv-ir}(a) includes $N_0(t)$ using a mask function width $W=1.5$ a.u..
It closely follows the XUV ionization yield when both curves are normalized to 1.
Varying the mask function width, the overall shapes of the electron density still follows the XUV ionization yield, but the perfect agreement between them is lost.
In a field free system the electron density at the origin is dominated
by the ground state population, also shown in Fig.~\ref{fig:xuv-ir}(a).
That quantity has similar characteristics as the XUV probe signal, but the
modulations are less pronounced, which indicates that excited and
continuum states influence the XUV probing process.
Note, however, that the effect of the excited state population after the pulse is rather small
compared to the modulations of the XUV yield during the pulse.
Therefore the participation of excited states during the ionization process can only
be of a transient nature, as in adiabatic distortions by the strong field\cite{xie12}.

From the electron density near the nucleus, we observed that there are fast oscillations at the end of the IR pulse.
To get insight into this observation, we zoom in this region and extend the time windows to 24 fs (NOTE: the IR pulse is over at 20 fs).
To show the agreement between the electron density near the nucleus and the XUV yields, we carry out simulations with denser time delay points.
Obviously, the perfect agreement preserves.
The fast oscillations probed by XUV-probing are similar to the XUV absorption signal measured in experiments\cite{uiber07}, which are induced by excitation during the IR field interaction with the atom
Some part of electron the wave function can be excited to n\textit{p} states from the ground 1\textit{s} state through dipole transition.
The quantum beating between the ground state and the excited states leads to the fast oscillations in the electron density.
To further confirm the observed excitation effect, we perform Fourier transformation on the fast oscillating signal in the electron density.
As shown in Fig.~\ref{fig:xuv-ir}(c), the beating frequencies between the 1\textit{s} ground state and the excited 2\textit{p}, 3\textit{p} and 4\textit{p} states are evidently presented.
Such results prove that the XUV probing can serve as a method to study the excitation dynamics of an atomic or molecular system with attosecond temporal resolution.

In cases of NIR-probing and XUV-probing, different quantities are probed in these two situations due to different ionization mechanisms.
Tunneling ionization is more sensitive to the electron density distribution near the tunneling barrier.
In other words, the additional ionization yield induced by the NIR pulse reflects the electron density of the polarized system near the tunneling barrier which is some distance away from the origin.
On the other hand, the XUV pulse probes the electron density of the system near the origin, which is not only sensitive to the polarization of the electron wave packet in the laser field but also to the quantum beating between bound states.


In conclusion, we theoretically investigate and compare the NIR-probing and XUV-probing of the interaction of an atom with a strong IR field.
We found that the NIR pulse probes the electron wave function near the tunneling barrier which represents the polarization of the system, while the XUV pulse probes the electron wave function near the nucleus of the atom which can serve for the study of not only polarization but also excitation dynamics from the electron density oscillation due to the quantum beating between different bound states.
The underlying physics is different ionization mechanisms in these two cases.
It allows acquiring electron dynamics information from different spatial regions in the electron wave function of the system by varying the wavelength of the probing pulse.
Such knowledge can shed light on ongoing and future probing experiments on atoms and molecules using IR sources to visualize the ultrafast electron motion manifesting the IR field induced polarization and excitation.

We thank Dr. Ralph Ernstorfer for fruitful discussions and Christopher Nicholson for improving English in the letter. This work is financed by the Austrian Science Fund (FWF) (No.P25615-N27) and the European Union project CRONOS (No.280879-2). Thanks to Vienna Scientific Cluster for providing computing resource under project No.70458.

\end{document}